\documentclass[11pt]{article}
\pdfoutput=1
\usepackage{graphicx}
\usepackage{epsfig}
\usepackage{amsfonts}
\usepackage{amscd}
\usepackage{latexsym}
\usepackage{amsmath,amssymb}
\usepackage{verbatim}
\usepackage{setspace}
\usepackage{color}
\usepackage{fancyhdr}
\usepackage{cite}
\usepackage{hyperref}
\usepackage{tensor}
\usepackage{xfrac}
\usepackage{comment}

\usepackage[textheight=9in, textwidth=6.5in, letterpaper]{geometry}
\def\half{{1\over 2}}
\numberwithin{equation}{section}

 \def\bz{{\bar z}}
 
\def\0{{(0)}}
\def\1{{(1)}}
\def\2{{(2)}}

\def\<{\langle }
\def\>{\rangle }

\def\bz{{\bar z}}

\def\sl{$SL(2,{\mathbb R})$}
\def\slr{$SL(2,{\mathbb R})_R$}
\def\slw{$SL(2,{\mathbb R})_{\rm w}$}
\def\Wi{W$_{1+\infty}$}
\def\wi{w$_{1+\infty}$}
\newcommand{\bea}{\begin{eqnarray}}
\newcommand{\eea}{\end{eqnarray}}
\newcommand{\be}{\begin{equation}}
\newcommand{\ee}{\end{equation}}
\newcommand{\ba}{\begin{aligned}}
\newcommand{\ea}{\end{aligned}}

\def\be{\begin{equation}}
\def\ee{\end{equation}}
\def\beq{\be\begin{array}{c}}
\def\eeq{\end{array}\ee}

   \makeatletter
  \let\over=\@@over \let\overwithdelims=\@@overwithdelims
  \let\atop=\@@atop \let\atopwithdelims=\@@atopwithdelims
  \let\above=\@@above \let\abovewithdelims=\@@abovewithdelims
\renewcommand\section{\@startsection {section}{1}{\z@}%
                                   {-3.5ex \@plus -1ex \@minus -.2ex}
                                   {2.3ex \@plus.2ex}%
                                   {\normalfont\large\bfseries}}

\renewcommand\subsection{\@startsection{subsection}{2}{\z@}%
                                     {-3.25ex\@plus -1ex \@minus -.2ex}%
                                     {1.5ex \@plus .2ex}%
                                     {\normalfont\bfseries}}

\usepackage{tikz}
\usepackage{slashed}
\usetikzlibrary{calc}   
\usetikzlibrary{topaths}
\usetikzlibrary{decorations}
\usetikzlibrary{decorations.pathmorphing}

{\cfoot{\thepage}}
\begin{document}
\onehalfspacing
\begin{titlepage}
\unitlength = 1mm~\\
\vskip 5cm
\begin{center}

{\LARGE{\textsc{\wi\ and the  Celestial Sphere}}}

\vspace{0.8cm}
Andrew Strominger \\
\vspace{1cm}

{\it  Center for the Fundamental Laws of Nature, Harvard University, Cambridge, MA 02138, USA} \\

\vspace{0.8cm}

\begin{abstract}

It is shown that the  infinite tower of tree-level plus-helicity soft graviton symmetries in asymptotically flat 4D quantum gravity can be organized into a single chiral 2D Kac-Moody symmetry based on the wedge algebra of \wi\ which naturally acts on the celestial sphere at null infinity.  The infinite towers of soft photon or gluon symmetries also transform irreducibly under  \wi.  
 \end{abstract}

\end{center}

\end{titlepage}

\tableofcontents
\section{Introduction}

A central problem in physics is to find all the fundamental nontrivial  symmetries of nature implied by all the experimentally verified physical laws. By this we mean symmetries associated via  Noether's therorem to conservation laws with observable consequences. A hundred years ago the answer to this question would have been a short list including Poincare symmetries, (which lead to energy-angular momentum conservation) plus  a few more.\footnote{We do not consider here the rich variety of  emergent symmetries also observable in nature.}

In the 1960s BMS \cite{Bondi:1962px,Sachs:1962wk} showed that the answer cannot  be so simple because there is no sense in which the diffeomorphism group of 
general relativity (GR) in asymptotically flat spacetimes can be reduced to the Poincare group. They did not however (due to then-uncertainties about the structure of asymptotic infinity) either  identify an alternate larger asymptotic symmetry group of the full past and future spacetime
or  associate any observable conservation laws. Recently this problem has been translated  into the language of quantum field theory and Feynman diagrams  where it becomes  equivalent to identifying  soft theorems. A soft theorem is a  linear relation between scattering amplitudes (in asymptotically past and future flat spacetimes) in which one particle becomes `soft" in that its energy is taken to zero. Such theorems can always be recast as a 
conservation-law-implying symmetry\cite{Strominger:2013lka,Strominger:2013jfa,He:2014laa}. Completing the program of  BMS, an infinite number of exact conservation laws were thereby discovered which relate arbitrary moments of ingoing and outgoing energy-momentum fluxes to measurable gravitational memory effects \cite{Strominger:2014pwa}. 

These developments were  satisfying but far from the end of the story. Soft theorems abound in both gauge theory and  gravity, with more being discovered only recently, and each associated to an infinite number of symmetries and measurable conservation laws.  Moreover, the known symmetries do not close under commutation, implying  an infinite tower of soft theorems \cite{Hamada_2018,Li_2018,Banerjee_2021,Pasterski:2021fjn}.

Hence despite all the progress, finding all the  symmetries and conservation laws implied by the standard model plus GR remains an outstanding open problem. 

In this letter we solve the problem in the limited context of the  tree-level  approximation with vanishing cosmological constant. Moreover we make the significant restriction to symmetries associated with plus-helicity soft particles only. 
Under these conditions we show that the soft symmetries can be succinctly described by a certain well-known infinite-dimensional $w$-symmetry group.

 We rely heavily on recent progress made in \cite{Guevara:2021abz}  for  tree-level gravity and gauge theory. In this work plus-helicity soft symmetries were compactly represented by 2D (higher-spin) currents in the celestial conformal field theory (CCFT)\footnote{This paper employs a bottom-up approach in which the CCFT is simply defined by a Mellin transform of  gauge and gravity scattering  amplitudes whose conformal properties are deduced from soft theorems. This differs from the top-down approach usually employed in the AdS/CFT correrspondence, where the boundary CFT can be  constructed from first principles using string theory. A top down approach is not available here as we strive to describe the real world for which the complete microscopic fundamental laws are unknown.}  living on the celestial sphere at null infinity. An infinite tower of such  currents and their algebra were derived at tree level  using  positive-helicity soft theorems and the celestial operator product expansion (OPE). 
  
 The somewhat lengthy results of \cite{Guevara:2021abz} were displayed in a basis with manifest covariance under the global Lorentz-conformal group of the celestial sphere.  The currents include an  `\slw ' Kac-Moody algebra arising from the subleading soft graviton theorem.  In this paper we find dramatic simplifications by reorganizing the generators according to representations of this \slw, which is accomplished with a  version of the light ray transform \cite{Kravchuk:2018htv}. The entire tower of currents is assembled into a chiral Kac-Moody symmetry of the wedge algebra of \wi!  This same Kac-Moody algebra has appeared previously including in the Penrose  twistor construction \cite{Penrose:1976js},
 in discrete states of the $c=1$ string \cite{Klebanov:1991hx} and  \wi-gravity\cite{Pope:1989sr}. Moreover we find, unlike in \cite{Guevara:2021abz}, in the \slw\  covariant  presentation only positive half-integral weights appear. 
  
 In the next section we present explicity the algebraic transformation from the conventional to the  \slw\ covariant basis, and show that the entire algebra is the Kac-Moody symmetry of the wedge algebra of \wi. In section 3 the infinite tower of soft symmetries for gauge theory are written in \slw\ covariant form and again found to dramatically  simplify. Again only positive \slw\ weights appear. These results suggest that \wi\ - or perhaps its quantization \Wi\ \cite{Pope:1989sr}- provides an organizing principle 
 for CCFT.  We conclude with speculations on this role in section 4. 

\section{Gravity}

Let us recap the basic results and notation of \cite{Guevara:2021abz} for gravity.   Let $G_{\Delta}^{+}(z,\bar{z})$
denote the positive-helicity conformal-primary graviton operator with 2D conformal weight $\Delta$ which crosses  the celestial sphere at a point $(z, \bz)$.  Define a discrete family of conformally soft\footnote{The conformally soft gravitons defined here differ from the usual energetically soft gravitons 
in that the conformal weight, rather than the enegy, is taken to a limiting value. Nevertheless they obey soft theorems which mirror their energetically-defined counterparts \cite{Pate:2019mfs}.}positive-helicity gravitons
\begin{equation} \label{eq:Hops}
H^{k} = \lim_{\varepsilon \to 0} \varepsilon G_{k + \varepsilon}^{+} , \ \ \ \ k = 2,1, 0, -1,  \ldots, 
\end{equation}
with weights
\begin{equation} \label{eq:Hweights}
 (h,\bar h)= \left(\frac{k+2}{2},  \frac{k-2}{2}\right),
\end{equation}
and a consistently-truncated antiholomorphic mode expansion\footnote{Outside the specified range of $n$, the \slr-invariant norm vanishes. Such operators may still have contact interactions but in this paper operators are always at distinct  points.}  
\begin{equation} \label{eq:Hmodes}
\begin{aligned}
  H^{k}(z,\bar{z}) &= \sum_{n = \frac{k-2}{2}}^{\frac{2-k}{2}} \frac{H^{k}_{n}(z)}{\bar{z}^{n + \frac{k-2}{2}}}.   \end{aligned}
\end{equation}
Each $H^k_n(z)$ is a 2D symmetry-generating conserved current whose Ward identity is a soft theorem. The factor of $\varepsilon$ in \eqref{eq:Hops} cancels the soft poles, allowing a finite OPE. Throughout this paper  $z$ and $\bar z$ are treated as independent, which amounts to continuing (3,1) Minkowski space to (2,2) Klein space, the celestial sphere to the celestial torus \cite{Atanasov:2021oyu} and Lorentz-$SL(2,\mathbb C)$ to $SL(2,\mathbb R)_L\otimes SL(2,\mathbb R)_R$.  The $n$-index in \eqref{eq:Hmodes} then transforms in the $3-k$ dimensional representation of the \slr.
The simplest example of  the $k=1$ term generates supertranslations. Expanding $H^1_{\pm \half}(z)=\sum_mH^1_{\pm \half, m}z^{-m-3/2}$, the four modes $H^1_{\pm \half,\pm \half}$ generate the four global translations. $k=0$ is related to superrotations and includes Lorentz transformations. 
Defining the commutator for holomorphic objects\footnote{Note that this is a 2D celestial commutator on a 1D circle, not to be mistaken for a 4D commutator on a 3D slice \cite{Crawley:2021ivb}.}
\be  \left[A,B\right](z) = \oint_z \frac{dw}{2\pi i} A(w)B(z),\ee
the soft current algebra found in \cite{Guevara:2021abz} for gravity is\footnote{We note that the $(2,0)$ current $H^2_0(z)$ commutes with all other generators and is a central term in the supertranslation algebra. It has been taken to vanish in most applications but is natural to include here.}
\begin{equation}\label{sxa}
\left[H^{k}_{m},H^{l}_{n}\right] = -\frac{\kappa}{2} \left[ n(2-k) - m(2-l) \right]\frac{(\frac{2-k}{2} -m + \frac{2-l}{2} - n -1)!}{( \frac{2-k}{2} - m)!(\frac{2-l}{2} - n)!} \frac{(\frac{2-k}{2} + m + \frac{2-l}{2} + n - 1)!}{(\frac{2-k}{2}+m)!(\frac{2-l}{2}+n)!} H^{k+l}_{m+n},
\end{equation}
 To write the algebra \eqref{sxa} in a simpler form we define
\be\label{hw}  w^p_n=  { 1 \over \kappa}(p-n-1)!(p+n-1)!H^{-2p+4}_n.\ee
This is essentially the light-transform or right-shadow \cite{Kravchuk:2018htv} in one dimension adapted to finite \sl\ representations, and $p$ is essentially the righ-shadowed weight.  
\eqref{sxa} then becomes
\be \label{scd} [w^p_m,w^q_n]=\left[m(q-1) -n(p-1)\right]w^{p+q-2}_{m+n}. \ee 
Since the index  $k$ on $H^k_m$ runs over $k=2,1,0,....$,  $p$ runs over the positive half integral values \be\label{dsc} p=1,{3 \over 2}, 2, {5 \over 2},....\ee
The restriction  ${k-2 \over 2}\le m \le {2-k \over 2}$ becomes\footnote{For example for the $p=2$ Virasoro case this restricts to the $SL(2,\mathbb R)$  current subalgebra.}
\be\label{rst}1-p \le m \le p-1.\ee 
This is one of our  main results. 

The commutators  \eqref{scd} were first written down  by Ioannis Bakas in 1989 \cite{Bakas:1989xu}. In this work 
$m$ is an arbitrary integer  and the resulting algebra is now referred to as \wi. The closed $p=2$ subalgebra of \wi\ is the $c=0$ Virasoro algebra. 
This algebra in \eqref{scd}  is a subgroup of   \wi\  with $m$ restricted according to \eqref{rst}. The  resulting restricted algebra is known as the wedge subalgebra of \wi, and can also be rewritten as $GL(\infty,{\mathbb R})$\cite{Pope:1989sr}.  

Moreover, each $w_m^p(z)$ acts as a current on the celestial sphere. Hence we have a $GL(\infty,{\mathbb R})$ Kac-Moody  algebra.  The modes of the Kac-Moody currents act on the states of the 2D CCFT associated by the state-operator correspondence to operator insertions in the celestial sphere \cite{Crawley:2021ivb}.  In fact precisely this Kac-Moody algebra has been studied previously both as  the symmetry group of $c=1$ string theory \cite{Klebanov:1991hx} and 2D  \wi-gravity \cite{Pope:1991zka}.\footnote{Here we take the classical limit \Wi\ $\to$\wi\ of the quantum symmetry \Wi.}
A review of this and other fascinating aspects of W algebras can be found in \cite{Pope:1991ig}.
We return to this  connection in the last section. 

The closed subgroup generated by \be L_m\equiv w_m^2,~~~m=-1,0,1\ee is an  `\slw'  current algebra
implied by the subleading soft graviton theorem. It is related to the original generators 
in \eqref{sxa} by 
\be L_0={1 \over \kappa} H^0_0,~~L_{\pm 1}={2\over \kappa }H^0_{\pm 1}.\ee
$w^q_n$ transforms under this \slw\ as 
\be [L_m,w^q_n]=\left[m(q-1) -n\right]w^{q}_{m+n}. \ee 
like the modes of an \slw\ primary operator of \slw\ weight  $q$. Since $q\ge 1$ here, these are all positive. However, unlike the $H^k_n$,  they do not transform canonically under the original \slr\ because of the mode-dependent relation between the two. Indeed the normalization of the  $H^k_n$ modes was chosen so that they transform canonically under the original \slr\ conformal generators with weights   $\bar{h} =   \frac{k-2}{2}$. 

Evidently \slr\ and \slw\ are not quite the  the same thing. $w_n^p$ on the LHS of \eqref{hw} lies in a  
positive  weight $h_w=p$ representation of \slw, while $H_n^{-2p+4}$ on the RHS lies in a $2p+1$ dimensional, 
negative  weight $\bar h =1-p$ representation of \slr. \eqref{hw} is the  relation between them. Both representations are $2p+1$ dimensional, because finite weight $h$ \sl\ representations are $2h-1$ dimensional for positive half integral $h$ and $-2h-1$ dimensional for negative half integral $h$.  

Significantly, the \slw\ representations which appear here are all positive weight.

\section{Gauge theory}
Now we turn to nonabelian gauge theory. Let ${O}_{k}^{a,+}(z,\bar{z})$ denote a positive helicity,  conformal weight $k$ gluon operator with adjoint group index $a$ at the point $(z, \bar z)$ on the celestial sphere. 
Mode-expanding in $\bar z$  on the right 
\begin{equation}
{O}_{k}^{a,+}(z,\bar{z}) = \sum_n\frac{{O}^{a,+}_{k,n}(z)}{\bar{z}^{n+{k-1 \over 2}}},
\end{equation}
 conformally soft currents are  defined by
 \begin{equation} \label{eq:Ropsx}
  R^{k,a}_n(z)
     := \lim_{\varepsilon \to 0} \varepsilon O_{k + \varepsilon,n}^{a,+}(z), ~~~~~ \ \ \ \ k = 1, 0, -1, -2, \ldots, ~~~~~~\quad {k-1 \over 2}\le n\le {1-k \over 2}.
\end{equation}
This has \sl$_L\otimes$\slr\ weights 
\begin{equation} \label{eq:Rweights}
  (h, \bar{h}) = \left(\frac{k+1}{2},  \frac{k-1}{2}\right). 
\end{equation}
These values of conformal weights $\Delta=k$  include all the conformally soft poles encountered in the OPE \cite{Fan:2019emx,Pate:2019lpp}.  The factor of $\varepsilon$ incorporated in \eqref{eq:Ropsx} is needed to cancel these poles, leading to finite OPEs for the rescaled $R^{k,a}$.

The soft current algebra for gauge theory  is \cite{Guevara:2021abz}
\be \left[R^{k,a}_{n},R^{l,b}_{n'}\right] = -i f^{ab}{}_c \frac{( \frac{1-k}{2} -n + \frac{1-l}{2} - n')!}{(\frac{1-k}{2} - n)!( \frac{1-l}{2} - n')!} \frac{(\frac{1-k}{2} + n + \frac{1-l}{2} +n')!}{(\frac{1-k}{2}+n)!(\frac{1-l}{2}+n')!} R^{k+l-1,c}_{n+n'}.\ee
Let us define
\be S^{q,a}_m=(q-m-1)!(q+m-1)!R^{3-2q,a}_{m},\ee
where $q=1,{3 \over 2},2,...$. 
One then finds the simple algebra
\be \left[S^{q,a}_{n},S^{p,b}_{n'}\right] = -i f^{ab}{}_c S^{q+p-1,c}_{n+n'}.\ee
Moreover using \cite{Guevara:2021abz}
\begin{equation} \label{eq:HRope}
\left[H^{k}_{m},R^{l,a}_{n}\right] = -\frac{\kappa}{2} \left[ n(2-k) - m(1-l) \right]\frac{(\frac{2-k}{2} -m + \frac{1-l}{2} - n - 1)!}{( \frac{2-k}{2} - m)!(\frac{1-l}{2} - n)!} \frac{(\frac{2-k}{2} + m + \frac{1-l}{2} + n- 1)!}{(\frac{2-k}{2}+m)!(\frac{1-l}{2}+n)!} R^{k+l,a}_{m+n},
\end{equation}
one finds the irreducible representation
\begin{equation} \label{eq:HRope}
\left[w^{p}_{m},S^{q,a}_{n}\right] =  \left[m(q-1)- n(p-1)  \right]S^{p+q-2,a}_{m+n}.
\end{equation}
The \slw\ transformation of $S^q_m$
\begin{equation} \label{eq:HRope}
\left[L_{m},S^{q,a}_{n}\right] =  \left[m(q-1)- n  \right]S^{q,a}_{m+n}.
\end{equation}
is that of a primary of weight $q$. Again we find only positive weights.

\section{Speculations}

The appearance of W-algebras in the 2D celestial symmetry group connects celestial holography to several other research areas. 
We find it irresistible to speculate on what might lie ahead as these connections unfold. 

\wi\ has a natural deformation to \Wi, a significantly more complicated algebra.  This deformation can be understood as arising from quantization.  In the context of specific classical 2D realizations of \wi, anomalies encountered in quantization deform the classical  \wi\ algebra to the quantum \Wi\cite{Bergshoeff:1991un}.  While  currently  not well-understood, the tree-level soft algebra of 4D quantum gravity also gets deformed on quantization, as implied among other things by one-loop corrections to the soft theorems. So it is natural to speculate that the action of \wi\ on this soft algebra is deformed to  \Wi\ in the 4D quantum theory of gravity. 

One might have expected a Virasoro algebra rather than an \slw\ current algebra among the symmetries. However the two are closely related in many similar contexts. For example  the Chern-Simon formulation of AdS$_3$ gravity, at first sight gives an \slr$\otimes SL(2,{\mathbb R})_L$ current algebra on the boundary. However, the AdS$_3$ boundary conditions implement a Hamiltonian reduction to Vir$_R \otimes$Vir$_L$ \cite{Banados:1994tn,Coussaert:1995zp}.\footnote{  A similar reduction could perhaps be operative here from constraints related to IR divergences.} The Virasoro generators are field-dependent \sl\ transformations which lie in the enveloping algebra of  the  \slr$\otimes SL(2,{\mathbb R})_L$ current algebra.  In another example, 2D gravity in light cone gauge exhibits an \slr\ current  algebra \cite{Polyakov:1987zb}, but the same theory in conformal gauge exhibits a Virasoro symmetry \cite{David:1988hj}, again indicating a relation between the two symmetry actions.   

This type of relation has an infinite-dimensional uplift to the present context. \sl\ is  the wedge algebra of Virasoro,  just as $GL(\infty,{\mathbb R})$ is the wedge algebra of \wi\cite{Pope:1989sr}.  Just as   Virasoro lies in the  enveloping algebra of \sl\ Kac-Moody, \wi\ (or \Wi\ at the quantum level) lies in the enveloping algebra of $GL(\infty,{\mathbb R})$ Kac-Moody 
 \cite{Pope:1991zka}. So it is natural to speculate that 
\Wi\ is a celestial symmetry of 4D quantum gravity.

\wi\ also appears as the symmetry group of classical self-dual gravity in (2,2) signature Klein space \cite{Boyer:1985aj}, whose sole degree of freedom is given by the Kahler potential. The self-duality suggests that the dual CCFT  should be chiral, and acted on (at bulk tree level) by \wi .  
This is a natural context in which to study celestial holography, because the complicated interactions of opposite chirality sectors and double soft-limit ambiguities are absent. The $N=2$ string describes the quantization of this theory \cite{Ooguri:1991fp}, where \wi\ is potentially deformed to \Wi.  Perhaps 
the holographic CCFT dual of  $N=2$ string theory  in 4D Klein space is given by a 2D quantum  \Wi -gravity  theory on the celestial torus. 

We leave these thoughts to future explorations. 

\section*{Acknowledgements}

I am  grateful to Ofer Aharony, Adam Ball, Alfredo Guevara, Mina Himwich, Sruthi Narayanan  and Monica Pate  for useful conversations.  This work was supported by DOE grant de-sc/0007870. 
 \bibliography{woperators.bib}
\bibliographystyle{utphys}

\end{document}